\documentclass[twocolumn,showpacs,preprintnumbers,amsmath,amssymb]{revtex4}

\begin{document}
\title{Comment on ``Quantum string seal is insecure''}
\author{Guang Ping He}
 \email{hegp@mail.sysu.edu.cn}
\affiliation{School of Physics \& Engineering and Advanced
Research Center, Sun Yat-sen University, Guangzhou 510275,
People's Republic of China}

\begin{abstract}
Recently an attack strategy was proposed by Chau [H. F. Chau,
quant-ph/0602099 v3], which was claimed to be able to break all
quantum string seal protocols, including the one proposed by He
[G. P. He, Int. J. Quant. Inform. 4, 677 (2006)]. Here it will be
shown that the information obtained in He's protocol by the attack
is trivial. Thus Chau's conclusion that all quantum string seals
are insecure is wrong. It will also be shown that some other
claims in Chau's paper are inaccurate either.
\end{abstract}

 \maketitle

\newpage

In a recent paper \cite{Insecure}, Chau claimed that all quantum
string seals are insecure. The core of Chau's attack strategy is
the measurement
\begin{equation}
Q_{i0}=aI+b\left| i\right\rangle \left\langle i\right| .  \label{measurement}
\end{equation}
(see Eq. (29) of that reference). It was claimed that with this
measurement, the attacker can obtain non-trivial information on
the sealed string while escapes the verifier's detection with at
least $50\%$ chance. However, the paper concentrated only on the
fidelity of the sealed state corresponding to the attacker's
measurement, without providing a detailed evaluation on the amount
of information obtained by the attacker. Here it will be shown
that for a class of quantum string seal protocols including the
one proposed by He \cite{String}, this amount of information is
only trivial. Therefore in contrast to Chau's claim, quantum
string seal can be unconditionally secure.

In fact, the general proof on why Chau's attack strategy fails had already
been well addressed in Ref. \cite{Security}. Briefly, consider a simple
model of imperfect quantum string seal, in which the sealed state for the
message $i^{\prime }$ is taken as
\begin{equation}
\left| \tilde{\psi}_{i^{\prime }}\right\rangle =\sum\limits_{j^{\prime
}}\lambda _{i^{\prime }j^{\prime }}\left| \psi _{j^{\prime }}\right\rangle
_{B},  \label{model}
\end{equation}
where the notation is the same as that in Eq. (1) of Ref. \cite{Insecure}.
Applying the measurement $Q_{i0}$ on it yields
\begin{equation}
Q_{i0}\left| \tilde{\psi}_{i^{\prime }}\right\rangle =(a+b)\lambda
_{i^{\prime }i}\left| i\right\rangle +a\sum_{j\neq i}\lambda _{i^{\prime
}j}\left| j\right\rangle .  \label{final}
\end{equation}
Thus the probability for the message $i^{\prime }$ to be decoded as $i$ by
the attacker is
\begin{eqnarray}
p_{i^{\prime }i} &=&a^{2}+(2ab+b^{2})\lambda _{i^{\prime }i}^{2}  \nonumber
\\
&=&\frac{1-\nu }{N}+\nu \lambda _{i^{\prime }i}^{2},
\end{eqnarray}
where $\nu $\ is defined by Eq. (12) of Ref. \cite{Insecure}. According to
Sec. IV of Ref. \cite{Insecure}, by fixing $\nu =1/2$, the attacker can
escapes the verifier's detection at least half of the time, so that all
quantum seals are claimed to be insecure. But in this case, the above
equation becomes
\begin{equation}
p_{i^{\prime }i}=\frac{1}{2N}+\frac{\lambda _{i^{\prime }i}^{2}}{2}.
\end{equation}
It means that any one of the $N$ possible choices of the message
$i^{\prime } $ has at least the probability $1/(2N)$ to be decoded
as message $i$, even if its content is completely irrelevant with
$i$. In other words, whenever the attacker obtains a message $i$
via the measurement strategy, there is at less a probability
$p_{i}=\sum_{i^{\prime }}1/(2N)=1/2$ that the original message can
be anything, i. e., the amount of information he obtained is zero.
Thus it can be seen that the attack strategy is useless. Though at
half of the time it can escape the verifier's detection, the
amount of information obtained on the sealed message is only
trivial. Therefore Chau's claim that all quantum seals are
insecure is wrong.

Now it will be shown that the protocol proposed in Ref. \cite{String} is
indeed such a secure quantum string seal. In this protocol, to seal a string
$i^{\prime }=i_{1}^{\prime }i_{2}^{\prime }...i_{m}^{\prime }...$ ($%
i_{m}^{\prime }\in \{0,1\}$), the sealed state is taken as $\left| \tilde{%
\psi}_{i^{\prime }}\right\rangle =\sum_{m}\otimes \left| \tilde{\psi}%
_{i_{m}^{\prime }}\right\rangle $\ where $\left| \tilde{\psi}_{i_{m}^{\prime
}}\right\rangle =\cos \theta _{m}\left| i_{m}^{\prime }\right\rangle +\sin
\theta _{m}\left| \overline{i_{m}^{\prime }}\right\rangle $. Thus by taking
\begin{equation}
\lambda _{i^{\prime }j^{\prime }}=\prod\limits_{m}f_{m}(\theta _{m}),
\end{equation}
where $f_{m}(\theta _{m})$\ is $\cos \theta _{m}$\ ($\sin \theta
_{m}$) if the $m$-th bit of the string $j^{\prime }$ equals to
(does not equal to) that of the string $i^{\prime }$, we can see
that the protocol belongs to the class of quantum string seal
described by Eq. (\ref{model}). Therefore as shown above, it
cannot be broken by Chau's attack strategy.

In Sec. IV of Ref. \cite{Insecure}, it was claimed that ``the major loophole
in He's proof of the security of his quantum string seal in Ref. \cite
{String} is that he incorrectly assumed that measuring all the qubits is the
only method to obtain a significant portion of information of the sealed
message''. But this is obviously incorrect. In the paragraph before Eq. (5)
of Ref. \cite{String}, it was clearly written that the general security
proof starts as follows. Let $H$ denotes the $2^{n}$\ dimensional Hilbert
space where the sealed state lives in, and $V$ denotes the space where the
final state lives in after the attacker performs certain POVMs. Note that no
restriction was ever put on $V$. $V$ can even equal to $H$ if the attacker's
POVMs do not contain any projection operator which will make the sealed
state collapse. Thus every possible case is covered by the security proof
following that paragraph. There is no such assumption as mentioned in Chau's
claim.

It was also claimed in the same section of Ref. \cite{Insecure}
that the analog of the attack strategy proposed in Ref.
\cite{Security} is not optimal. In this analog, the attacker needs
no quantum computer to perform the collective measurement in Eq.
(\ref{measurement}). He can simply toss a coin to decide his
action. At half of the cases he performs the honest measurement
suggested by the quantum string seal protocol and reads the
string, while at the other half of the cases he does nothing. This
is completely equivalent to the $\nu =1/2$\ case of Chau's attack
strategy, because substituting $\nu =1/2$\ into Eq.
(\ref{measurement}) gives
\begin{equation}
Q_{i0}=\sqrt{\frac{1}{2N}}I+(\sqrt{\frac{1}{2}+\frac{1}{2N}}-\sqrt{\frac{1}{%
2N}})\left| i\right\rangle \left\langle i\right| .
\end{equation}
Due to the linearity of quantum mechanics, we can see that applying the
measurement $Q_{i0}$ ($i=0,...,N-1$) on the sealed state is equivalent to
applying the identity operator $I$\ (which actually means doing nothing)
with the probability $1/2$. The merit of the analog is that it can help us
understand clearly why Chau's attack can escape the verifier's detection at
half of the cases -- simply because the attacker has done nothing at these
cases. More generally, by tossing a biased coin, the attacker can have a
corresponding analog of Chau's strategy for any $\nu $\ value. Therefore
Chau's claiming that the analog of the attack strategy is not optimal sounds
confusing. It seems to indicate that the optimal strategy should have $\nu
=1 $\ instead of $\nu =1/2$. If so, Eq. (\ref{measurement}) becomes
\begin{equation}
Q_{i0}=\left| i\right\rangle \left\langle i\right| .
\end{equation}
Then Eq. (\ref{final}) shows that after applying $Q_{i0}$ on $\left| \tilde{%
\psi}_{i^{\prime }}\right\rangle $, the final state will collapse to $\left|
i\right\rangle $ with the probability $\lambda _{i^{\prime }i}^{2}$. Thus
the average fidelity of the final state is $\sum_{i}\lambda _{i^{\prime
}i}^{4}$, which is arbitrarily small as $N$ increases. Therefore it cannot
escape the verifier's detection. That is, the results in Ref. \cite{Insecure}
corresponding to different $\nu $\ values in fact shows that if the amount
of information obtained by the attack measurement is optimized, the
probability of escaping the detection will be trivial, or vice versa. In
either case, Chau's strategy is not a successful attack.

In addition, there is also a misleading claim in the introduction of Ref.
\cite{Insecure} (which also appeared in Ref. \cite{Chau}). It was claimed
that the security bounds of imperfect quantum single bit seal obtained by He
\cite{He} are not tight, while Chau proved that all imperfect quantum bit
seals are insecure, and obtained a greater lower bound \cite{Chau}. But in
fact, Chau's model of quantum bit seal studied in Ref. \cite{Chau} is less
general than that of He's in Ref. \cite{He}, and Chau's bound is not
tighter. More rigorously, in He's model, measuring the sealed states can
result in three outcome sets $G_{0}$, $G_{1}$\ and $\{g\notin G_{0}\cup
G_{1}\}$, where $G_{0}$ and $G_{1}$\ are corresponding to the decoded bit
values $0$ and $1$ respectively, while $\{g\notin G_{0}\cup G_{1}\}$ tells
the reader that the decoding fails \cite{He}. Also, the maximum probability $%
\alpha $ for the sealed bit $b$ to be read correctly can be kept
secret from the reader. Let $\beta $ denotes the probability for
the reading operation to be detected by the verifier. By proposing
an explicit cheating strategy, two security bounds $\beta
\leqslant 1/2$\ and $\alpha +\beta \leqslant 9/8$\ were obtained
in Ref. \cite{He}. But in Ref. \cite{Chau}, Chau's model covers a
special case of He's model only, where $\{g\notin G_{0}\cup
G_{1}\}=\emptyset $ and $\alpha $ (denoted as $q_{\max }$\ in that
reference) is known to the reader (otherwise his cheating
measurement cannot be constructed). The lower bound for the
fidelity of the resultant state (equivalent to $1-\beta $) was
also found, which was said to be greater than $1/2$. But in fact,
the greater lower bound is achieved only when the amount of
information obtained by the cheater drops. From the analog of the
attack strategy proposed in Ref. \cite {Security} it can easily be
seen that this result is not significant, because if the cheater
reads the sealed bit only with a small probability, the fidelity
of the resultant state is surely greater. Also, the result is in
agreement with $\beta \leqslant 1/2$, while no analog to the
finding $\alpha +\beta \leqslant 9/8$\ of Ref. \cite{He} was found
in Ref. \cite{Chau}. For this reason, the remark on Refs.
\cite{He} in Ref. \cite{Insecure,Chau} is improper.

The author would like to thank Helle Bechmann-Pasquinucci for valuable
discussions. This work was supported in part by the NNSF of China under
Grant No.10605041, the NSF of Guangdong province under Grant No.06023145,
and the Foundation of Zhongshan University Advanced Research Center.

\end{document}